\documentclass[a4paper,fleqn,usenatbib]{mnras}

\usepackage[T1]{fontenc}
\usepackage{ae,aecompl}

\usepackage{graphicx}	
\usepackage{amsmath}	
\usepackage{amssymb}	
\usepackage{txfonts}

\usepackage{color}
\usepackage{natbib}
\bibpunct[; ]{(}{)}{,}{a}{}{,}
\def\asec{\ifmmode ^{\prime\prime}\else$^{\prime\prime}$\fi}

\def\msun{M$_{\odot}$}

\def\degs{\ifmmode ^{\circ}\else$^{\circ}$\fi}
\def\amin{\ifmmode ^{\prime}\else$^{\prime}$\fi}
\def\asec{\ifmmode ^{\prime\prime}\else$^{\prime\prime}$\fi}

\def\psr{PSR~J2302+4442}
\def\degs{\ifmmode ^{\circ}\else$^{\circ}$\fi}
\def\amin{\ifmmode ^{\prime}\else$^{\prime}$\fi}

\def\eqalign#1{\null\,\vcenter{\openup1\jot \m@th
   \ialign{\strut\hfil$\displaystyle{##}$&$\displaystyle{{}##}$\hfil
   \crcr#1\crcr}}\,}
\title[The optical companion to PSR J2302+4442]{Optical identification of the binary companion to the millisecond PSR J2302+4442 with the Gran Telescopio Canarias}

\author[A. Yu. Kirichenko et al.]{
A. Yu. Kirichenko,$^{1,2}$\thanks{E-mail: aida.taylor@gmail.com}
S. V. Zharikov,$^{1}$
D. A. Zyuzin,$^{2}$
Yu. A. Shibanov,$^{2}$
A. V. Karpova,$^{2}$
\newauthor
S. Dai,$^{3}$
and A. Cabrera Lavers$^{4,5}$
\\
$^{1}$ Instituto de Astronom\'ia, Universidad Nacional Aut\'onoma de M\'exico, Apdo. Postal 877, Ensenada, Baja California, M\'exico, 22800\\
$^{2}$ Ioffe Institute, 26 Politekhnicheskaya st., St. Petersburg 194021, Russia\\
$^{3}$ CSIRO Astronomy and Space Science, Australia Telescope National Facility, Box 76, Epping, NSW 1710, Australia\\
$^{4}$ Instituto de Astrof\'isica de Canarias, 38200 La Laguna, Tenerife, Spain\\
$^{5}$ Departamento de Astrof\'isica, Universidad de La Laguna, 38206 La Laguna, Tenerife, Spain\\
}

\date{Accepted XXX. Received YYY; in original form ZZZ}

\pubyear{2018}

\begin{document}
\pagerange{\pageref{firstpage}--\pageref{lastpage}}
\maketitle

\begin{abstract}

We report detection of the binary companion to the millisecond pulsar J2302+4442 based on the deep observations performed with the Gran Telescopio Canarias.
The observations revealed an optical source with $r'$ = 23.33$\pm$0.02 and $i'$=23.08$\pm$0.02, whose position coincides with the pulsar radio position. 
By comparing the source colour and magnitudes with the white dwarf cooling predictions, we found that it likely represents a He or CO-core white dwarf and  
estimated its mass of 0.52$^{+0.25}_{-0.19}$ \msun~and effective temperature of 6300$^{+1000}_{-800}$ K. 
Combining our results with the radio timing measurements, 
we set constraints on the binary system inclination angle $i=73^{+6}_{-5}$ degrees and the pulsar mass $m_p=3.1^{+2.7}_{-2.0}$ M$_{\odot}$. \\

\end{abstract}




\section{Introduction}


Among $\sim$~2600 pulsars currently listed in the ATNF Pulsar Catalogue\footnote{www.atnf.csiro.au/people/pulsar/psrcat/, \citet{manchester}}, 
there are over 300 pulsars with very short rotational periods ($P < $ 30 ms). These are so-called millisecond pulsars (MSPs) that are believed  
to have been spun up through the recycling process, i.e. through angular momentum transfer 
by accretion from main-sequence companions during their low and intermediate X-ray binary stages \citep{kogan,alpar}.  
The majority of the known MSPs are observed in binary systems, and, depending on the initial conditions, the nature of the companion star can be diverse. 
In most of the cases, however, the companion represents a He white dwarf (WD) (see, e.g., \citet{tauris2011}). 

Observations of binary MSP systems allow for measurements of their fundamental parameters such as masses of both pulsar and companion star. In some cases 
this can be achieved through radio timing measurements of the Shapiro delay \citep{shapiro}. However, such measurements can be significantly hampered in case of long MSP system 
orbital periods, extending the observational time spans to decades. In this case, optical observations can be used for independent measurements of the WD companion mass 
by comparing the photometric and/or spectroscopic results with the WD evolutionary tracks. On the other hand, the optical detectability highly depends on the distance to 
a particular MSP system. 
 
In this respect, a significant contribution to the MSP population was made by the \textit{Fermi} $\gamma$-ray telescope detections 
and the follow-up radio searches \citep{ray}. 
To date, $\gamma$-ray MSPs represent over 40 per cent of more than 200 $\gamma$-ray 
pulsars detected with \textit{Fermi} \citep{guillemot}, and about 20 per cent of all known MSPs\footnote{https://fermi.gsfc.nasa.gov/science/eteu/pulsars/}.  
The advantage of the \textit{Fermi} pulsars is that most of them are nearby objects \citep{saz} and their faint optical binary companions are expected  
to be easily detected with large-aperture optical telescopes.  

The millisecond PSR J2302+4442 ($P_s=5.19~\mathrm{ms}$, $\dot{P} = 13.9\times 10^{-21}$) has been discovered in the Nan\c{c}ay Radio Telescope follow-up observations 
of unidentified \textit{Fermi} $\gamma$-ray sources and later studied during the timing campaign with the 
Nan\c{c}ay, Jodrell Bank and Green Bank telescopes \citep{cognard}.
Table~\ref{table1} summarises the parameters of the pulsar system. Following the discovery in the radio, the $\gamma$-ray pulsations were also 
found in the \textit{Fermi} data.
In addition, a tentative pulsar X-ray counterpart with the 0.5--8 keV flux of $\sim$~4$\times10^{-14}$ erg cm$^{-2}$ s$^{-1}$ 
was found in the \textit{XMM-Newton} observations of the \textit{Fermi} field conducted prior to the radio pulsar identification \citep{cognard}. 

Using a nine-year dataset from the North American Nanohertz Observatory for Gravitational Waves, \citet{fonseca} have tentatively detected a Shapiro 
timing delay in the PSR J2302+4442 system. However, given the long orbital period $P_b$, only a small fraction of the Shapiro-delay signal was sampled and only a  
rough estimate on the pulsar companion mass $m_c=2.3^{+1.7}_{-1.3}$ M$_{\odot}$ was provided. The nature of the companion star thus remains unclear.

\begin{table}
\begin{center}
\caption{Parameters of the PSR J2302$+$4442 system. The pulsar period and age are adopted from \citet{cognard} and the period 
derivative is from \citet{arzoumanian}. 
The pulsar coordinates for the MJD 56279 epoch  
and the proper motion values are from \citet{matthews} and \citet{guillemot}, respectively. 
The system orbital parameters are taken from \citet{fonseca}.}
\label{table1}
\begin{tabular}{lccc}
\hline
\hline
\multicolumn{4}{c}{Pulsar timing and astrometric parameters}  \\
\hline
$P_s$ (ms)                                  &   \multicolumn{3}{c}{5.192324646411(7)}                          \\
$\dot{P}$                                   &   \multicolumn{3}{c}{13.9$\times 10^{-21}$}                        \\
$\tau$ (yr)                              &   \multicolumn{3}{c}{6.2$\times10^{9}$}                                           \\
$\alpha_{\rm{J2000}}$                              & \multicolumn{3}{c}{23$^{\rm{h}}$02$^{\rm{m}}$46$^{\rm{s}}$.97878(3)}   \\
$\delta_{\rm{J2000}}$                              & \multicolumn{3}{c}{$+$44$^{\circ}$42$^{'}$22.{\arcsec}0928(3)}          \\
$l$ (deg)  &   \multicolumn{3}{c}{103.40}                 \\
$b$ (deg)   &     \multicolumn{3}{c}{$-$14.00}                       \\
$\mu_{\alpha}$ (mas\,y$^{-1}$)                                    & \multicolumn{3}{c}{-0.05(13)}                             \\
$\mu_{\delta}$ (mas\,y$^{-1}$)                                    & \multicolumn{3}{c}{-5.85(12)}                           \\
\hline
\multicolumn{4}{c}{Keplerian elements}   \\
\hline
$P_b$ (days)                                      & \multicolumn{3}{c}{125.93529697(13)}                           \\
$x$ (lt-s)                                               & \multicolumn{3}{c}{51.4299676(5)}                                 \\
$i$ (deg)                                               & \multicolumn{3}{c}{54$^{+12}_{-7}$ }                             \\
$e$                                                & \multicolumn{3}{c}{0.000503021(17) }                            \\
\hline
\end{tabular}
\end{center}
\end{table}

According to the NE2001 electron-density model \citep{cordes}, 
the pulsar dispersion measure DM~$=13.762\pm0.006$ pc cm$^{-3}$ and the line of sight correspond to the distance of 1.18 kpc. 
The YMW16 model \citep{yao} assigns them a smaller distance of 0.86 kpc. Both estimates suggest the PSR J2302+4442 
system as a promising target for optical observations. Based on the \textit{Swift}'s Ultraviolet/Optical Telescope observations, 
only shallow optical and ultraviolet upper limits on the companion flux 
were established \citep{cognard}.

Here we present deep optical observations of the PSR J2302+4442 system carried out with the Gran Telescopio Canarias (GTC) and report a 
likely identification of the pulsar binary companion. The details of observations and data reduction are presented in Sect.~\ref{sec:obs}, 
the results and analysis are described in Sect.~\ref{sec:res} and discussed in Sect.~\ref{sec:sum}.

\section{Observations and data reduction}

\label{sec:obs}

\begin{figure*}
\setlength{\unitlength}{1mm}
\begin{center}
\begin{picture}(125,40)(0,0)
\put (0,-20) {
\includegraphics[width=62mm]{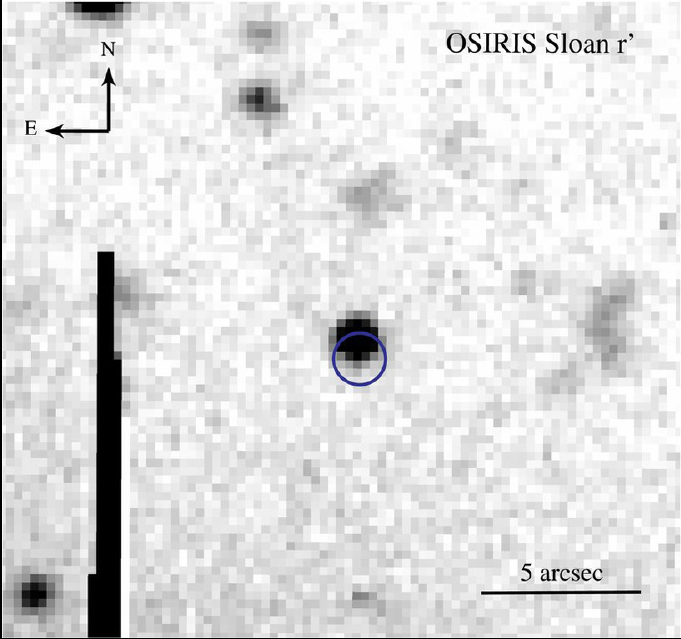}}
\put (80,-20) {
\includegraphics[width=62mm]{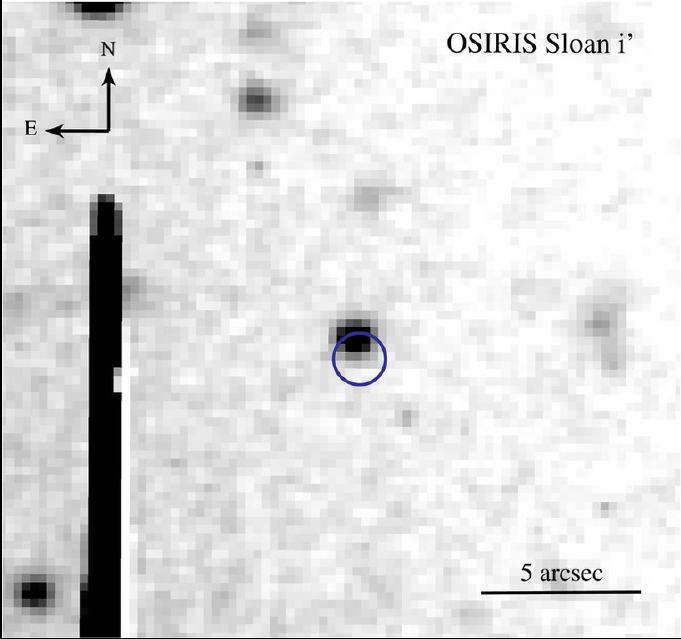}}
\end{picture}
\end{center}
\vspace{18mm}
\caption{GTC/OSIRIS Sloan 
$r'$ (left) and $i'$-band (right) image fragments of the \psr~field. 
The blue circle shows 3$\sigma$ radio timing pulsar position uncertainty
for the epoch of our observations (see text for details). The black stripe on both images is a part of a spike of a saturated field star.}
\label{fig:1}
\end{figure*}

The pulsar field was observed in the Sloan $r'$ and $i'$ bands with the Optical System for
Imaging and low-intermediate Resolution Integrated Spectroscopy
(OSIRIS\footnote{For instrument features see {http://www.gtc.iac.es/instruments/osiris/}})
at the GTC 
on September 16, 2017\footnote{Proposal GTCMULTIPLE2A-17BMEX, PI A. Kirichenko}. 
The detector provides 
a plate scale of 0.254 arcsec/pixel (2$\times$2 binning) and a field of view (FOV)
of 7.8 arcmin $\times$~7.8 arcmin consisting of two CCDs. The target source was exposed on CCD2. 
The observations were performed in dark time and under clear conditions, with seeing values in the range of 0.6$-$0.9 arcsec.
To avoid possible affection by bad pixels, we used 5 arcsec dithering between the individual exposures in both bands. 

We reduced the data using standard routines from the Image Reduction and Analysis Facility ({\tt IRAF}) package.
The individual images were bias-subtracted and flat-fielded. To stack the exposures  
together, we used 10 unsaturated field stars and the best-quality image as a reference in each band.  
The total integration times on the resulting combined $r'$ and $i'$-band images were 3135~s and 2415 s with the mean airmass values of $\approx$1.25 
and $\approx$1.13, respectively.

The astrometric solution was computed using a selection of 11 isolated non-saturated field stars on the resulting $r'$ and $i'$-images and their coordinates 
from the USNO-B1 astrometric catalogue. Formal $rms$
uncertainties of the astrometric fit were $\Delta$RA~$\la$ 0.098 arcsec 
and $\Delta$Dec~$\la$ 0.175 arcsec for both the $r'$ and $i'$-band images. 
The fit residuals were consistent with the nominal catalogue uncertainty of 0.2 arcsec. 
The resulting conservative 1$\sigma$ 
referencing uncertainty for the two images is
$\la$~0.22 arcsec for RA and $\la$~0.27 arcsec for Dec.

As a photometric reference, the SA 110-232 Sloan standard \citep{smith} was observed in both bands the same night as our target. 
Using the measured magnitudes and the mean OSIRIS extinction coefficients  
$k_{r'}$=0.07$\pm0.01$ and $k_{i'}$=0.04$\pm0.01$, we determined the magnitude
zero-points of 28.69(1) and 28.24(1) for the $r'$ and $i'$ bands, respectively.

\section{Results}
\label{sec:res}

\begin{figure}
\begin{center}
\begin{picture}(125,180)
\put (-80,0) {
\includegraphics[width=99mm]{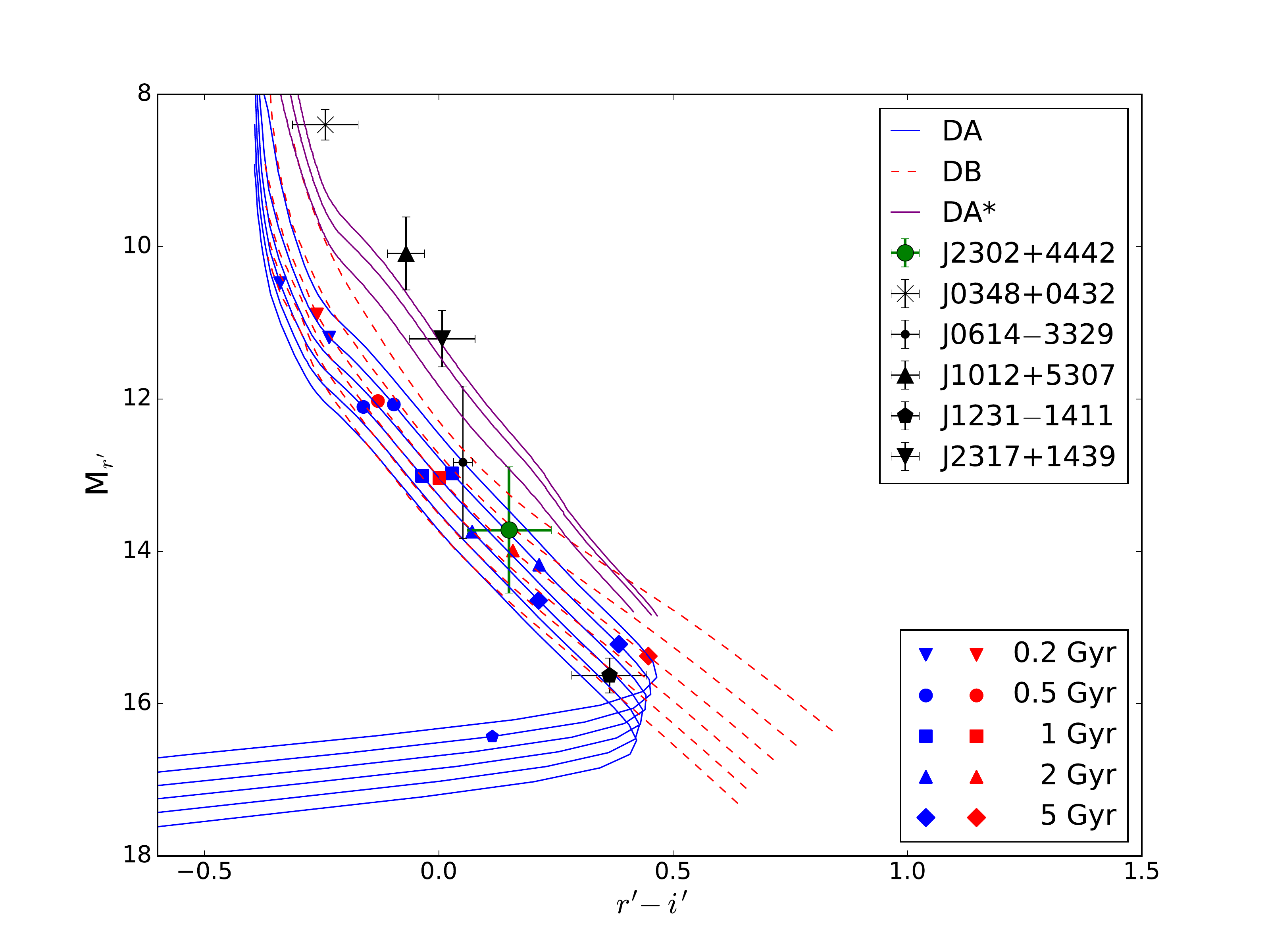}}
\end{picture}
\end{center}
\caption{Colour-magnitude diagram with the available data for companions to 
different MSPs including \psr~and the WD cooling tracks. 
Solid lines 
represent model predictions for WDs with hydrogen atmospheres with masses 0.1869, 0.2026 and 0.2495 ${\rm M_\odot}$ 
from \citet{panei} (purple, labelled as DA*) 
and 0.3--0.8 M$_{\odot}$ (spaced at 0.1 ${\rm M_\odot}$) from \citet{holberg}, \citet{kowalski}, \citet{tremblay} and \citet{bergeron2011} (blue, labelled as DA).  
Red dashed lines show tracks for WDs with helium atmospheres with masses 0.2--0.7 M$_{\odot}$ (spaced at 0.1 ${\rm M_\odot}$) 
from \citet{bergeron2011} (labelled as DB). Masses increase from upper to lower curves. 
The cooling ages are marked along the tracks.}
\label{fig:2}
\end{figure}

\begin{figure}
\begin{center}
\begin{picture}(125,180)
\put (-80,0) {
\includegraphics[width=99mm]{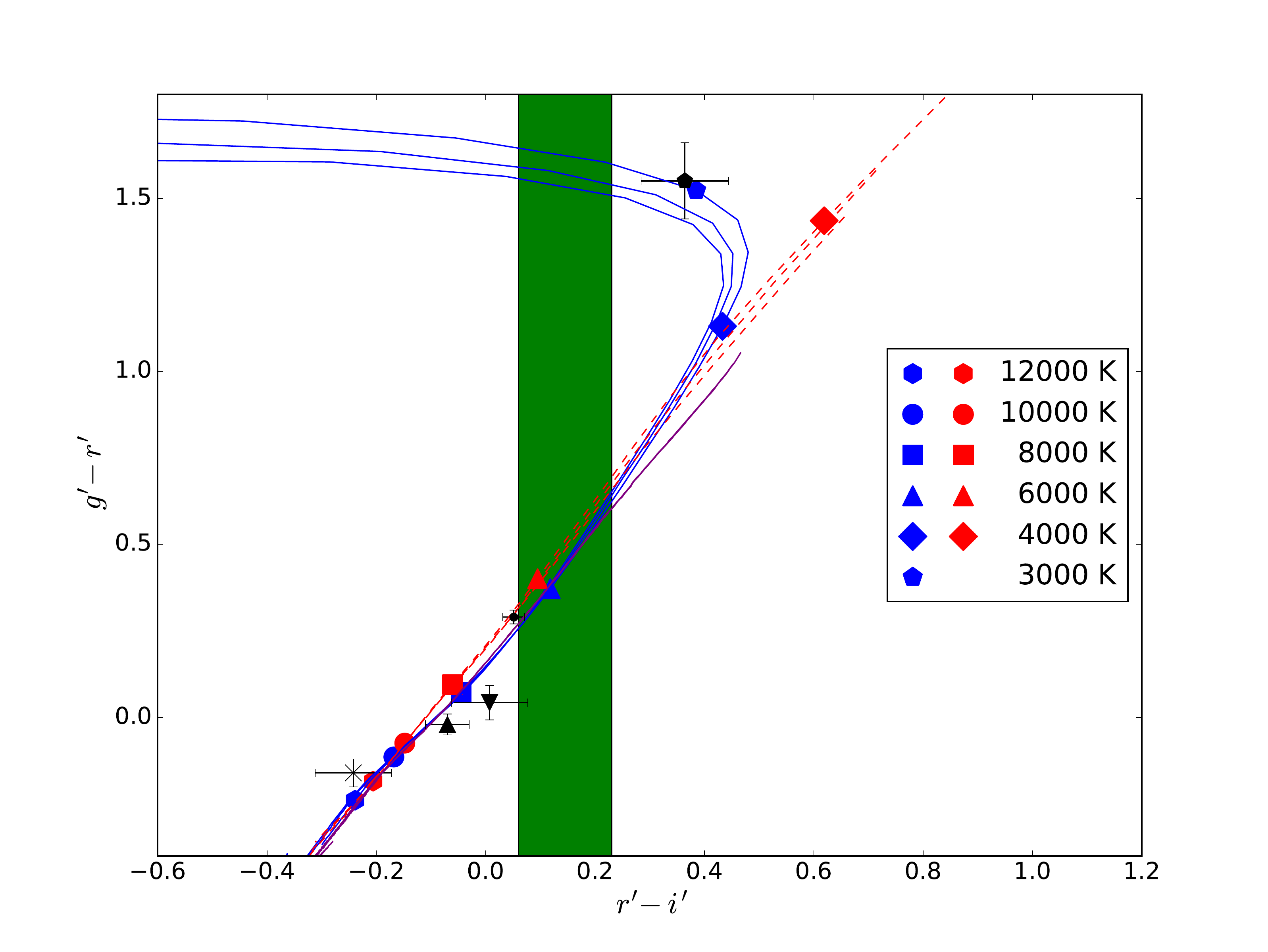}}
\end{picture}
\end{center}
\caption{Colour-colour diagram with the model predictions and the available data for the companions to 
MSPs demonstrated in Fig.~\ref{fig:2}, shown by the same colours and markers. The green stripe 
indicates the location range of the \psr~companion in the absence of the $g'$-band data. The effective temperatures are marked along the tracks.}
\label{fig:3}
\end{figure}

\begin{figure}
\setlength{\unitlength}{1mm}
\begin{center}
\begin{picture}(40,160)(0,0)
\put (-20,98) {
\includegraphics[width=78mm]{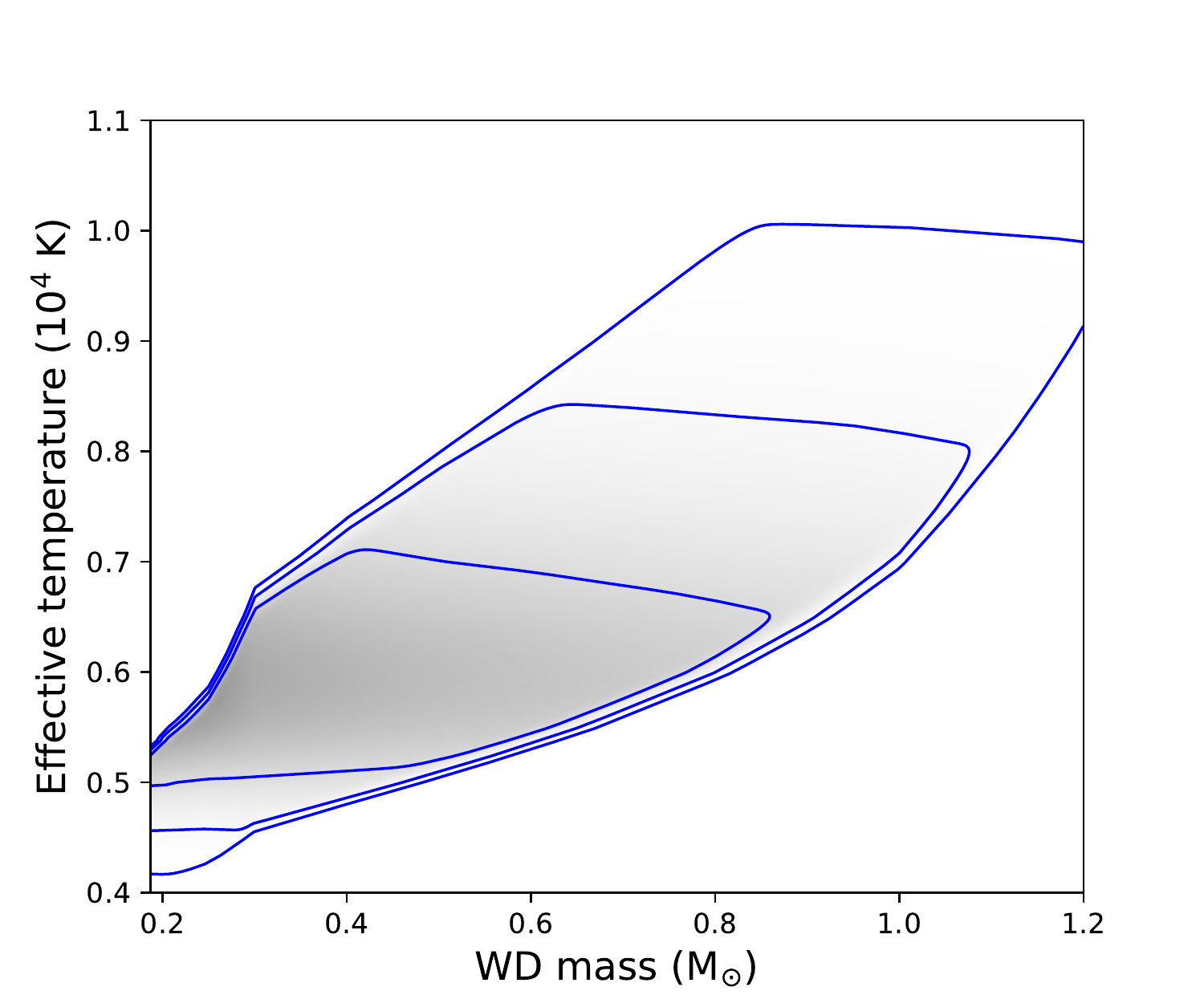}}
\put (-20,39) {
\includegraphics[width=78mm]{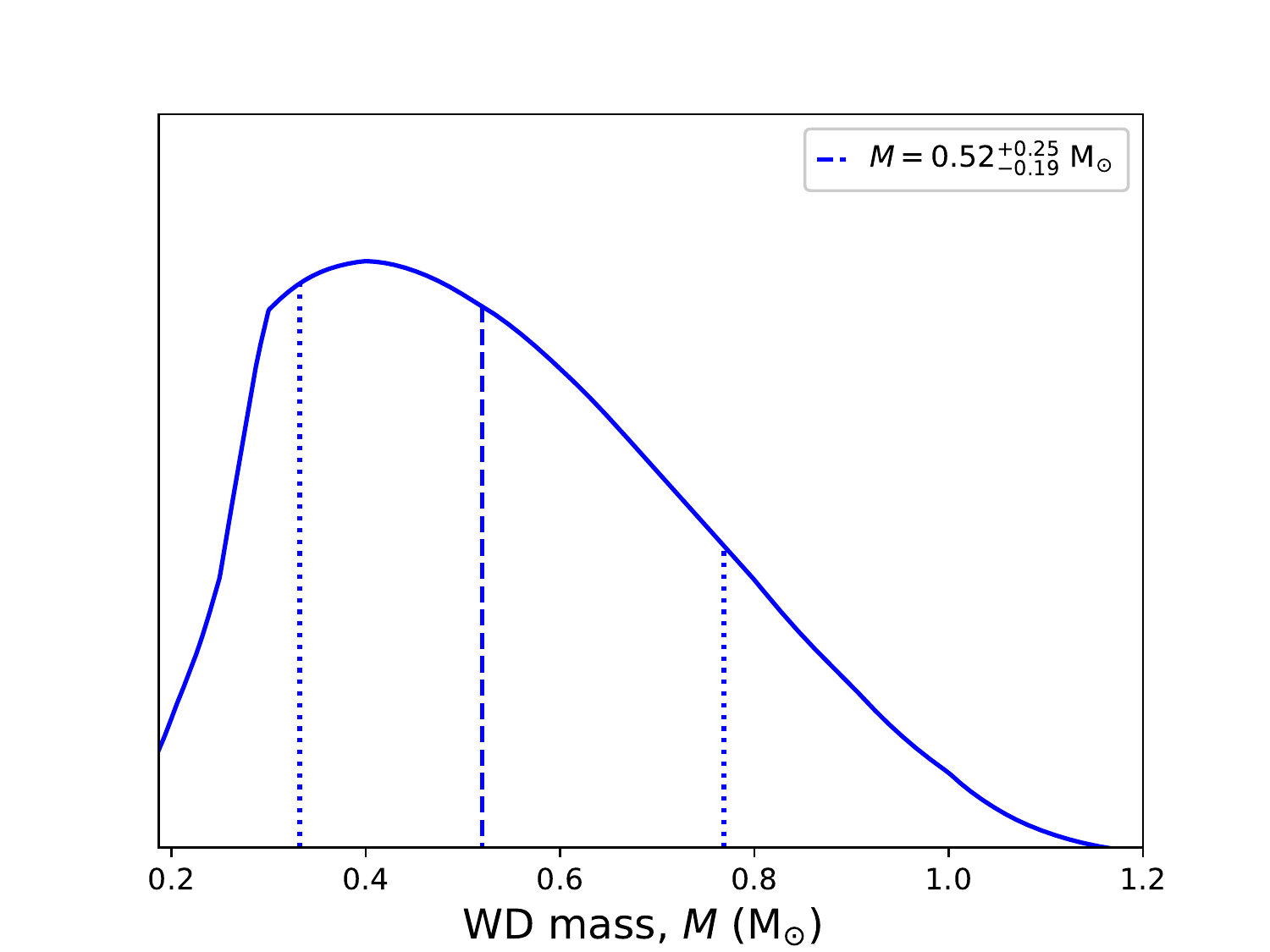}}
\put (-20,-20) {
\includegraphics[width=78mm]{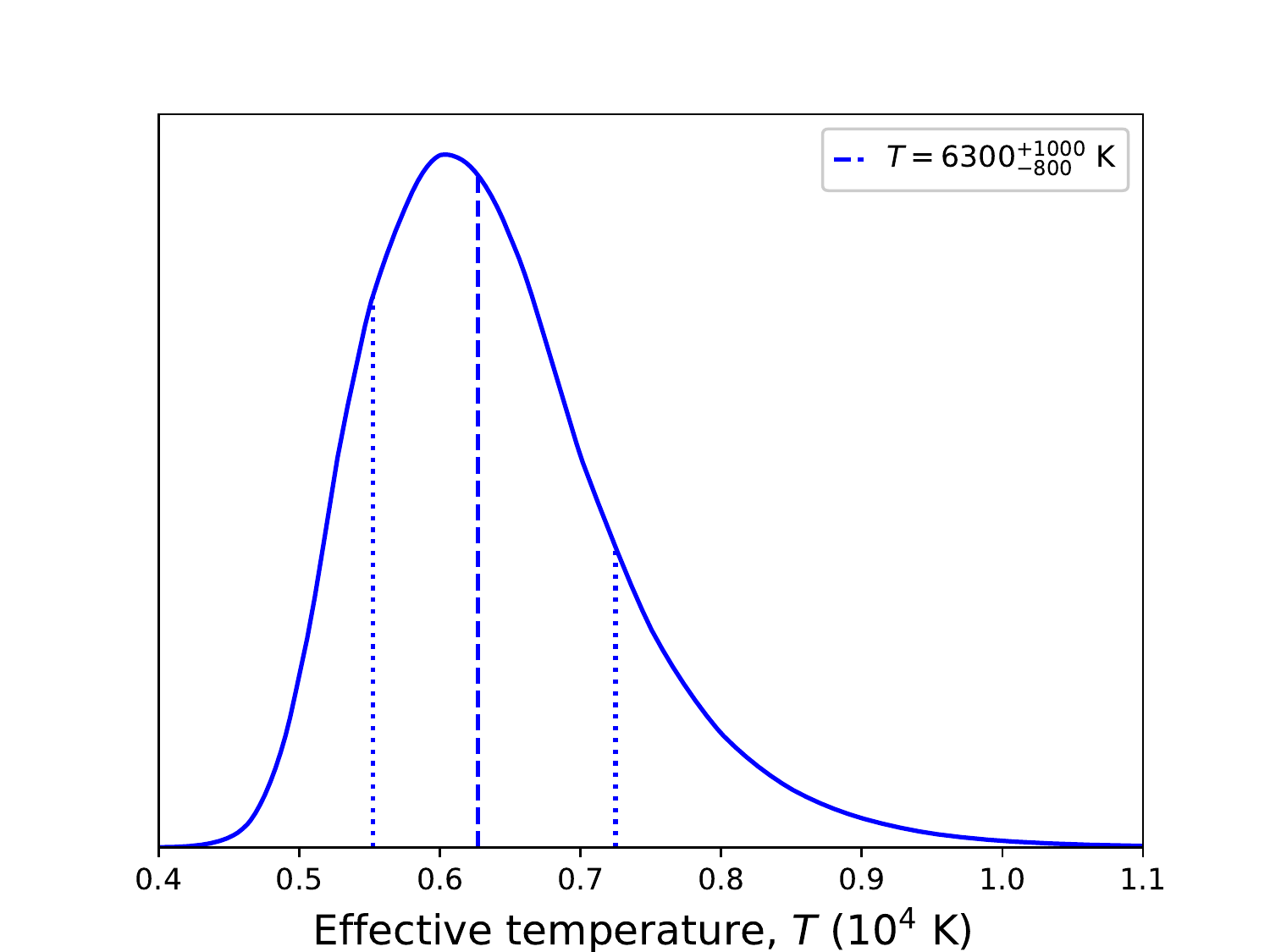}}
\end{picture}
\end{center}
\vspace{18mm}
\caption{Upper panel: constraints on the WD mass and effective temperature. The blue contours represent 1$\sigma$, 2$\sigma$ and 3$\sigma$ confidence levels. Middle 
and lower panels: the 1D likelihoods for the WD mass and effective temperature. Dashed and dotted lines correspond to the median value and 1$\sigma$ confidence levels, 
respectively.}
\label{fig:4}
\end{figure}

The fragments of the resulting $r'$ and $i'$-band images are presented in Fig.~\ref{fig:1}. The centre of the circle corresponds to    
RA~= 23$^{\rm{h}}$02$^{\rm{m}}$46$^{\rm{s}}$.979 and Dec~= $+$44$^{\circ}$42$^{'}$22.{\arcsec}06. 
The coordinates represent  
the pulsar radio timing position from \citet{matthews} shifted in accordance with the pulsar proper motion reported by \citet{guillemot}. 
The circle radius of 0.81 arcsec corresponds to
the 3$\sigma$ pulsar position uncertainty that accounts for the optical astrometric referencing and proper motion uncertainties.    
In both $r'$ and $i'$-band images we firmly detect a starlike source whose position with RA~= 23$^{\rm{h}}$02$^{\rm{m}}$46$^{\rm{s}}$.989 and 
Dec~= $+$44$^{\circ}$42$^{'}$22.{\arcsec}69 
falls into the pulsar position uncertainty region. Based on the spatial coincidence, we consider that this source is related to the pulsar system. 
The probability to detect an unrelated object within the pulsar positional 
region can be derived as $P$ = 1$-\exp(-\pi \sigma R^2)$, where $\sigma$ corresponds to the surface number density of stars with a similar magnitude and $R$ is the astrometric 
accuracy. Considering an unrelated object with a magnitude of 19$-$25, in case of our FOV 
this probability is as low as $\approx$0.002. 
Hence, we conclude that the source is a likely optical counterpart 
to the pulsar binary companion. We will hereafter refer to this object as the pulsar companion. 
Based on the aperture photometry performed on the resulting combined images, we estimated the companion magnitudes of $r'$ = 23.33(2) mag and $i'$ = 23.08(2) mag. 

The pulsar parallax is poorly constrained \citep{matthews, arzoumanian}, therefore, for the following estimations we will 
use the 95 per cent confidence 
lower limit on the pulsar distance of 0.5 kpc that follows from the parallax measurements (see \citet{arzoumanian}) and the upper limit of 1.0 kpc. The latter was 
estimated assuming the efficiency of conversion of spin-down power into $\gamma$-ray radiation, $\eta = L_{\gamma} / \dot{E}$, to be 100 per cent.         
 
Using the interstellar dust reddening model by \citet{green} and considering the pulsar distance range of 0.5--1.0 kpc, 
we obtained the reddening value E(B-V) = 0.16$\pm$0.03. Along with the conversion coefficients\footnote{The coefficients correspond to the $R_V$ = 3.1 reddening law} 
provided by \citet{schlafly}, this leads to the extinction correction values $A_{r'}$ = 0.37$\pm$0.07 and $A_{i'}$ = 0.27$\pm$0.05. The corresponding  
companion dereddened magnitues are $r'$ = 22.96$\pm$0.07 mag and $i'$ = 22.81$\pm$0.05 mag. 
 
Aiming to check whether the PSR J2302+4442 companion, as in the most probable case, belongs to the WD population,  
we compared its $r'$-band magnitude against the WD cooling predictions  
from \citet{holberg}, \citet{kowalski}, \citet{tremblay}, \citet{bergeron2011} and \citet{panei}. 
The respective colour-magnitude diagram with absolute magnitudes is presented in Fig.~\ref{fig:2}. 
The parameters of companions to some other millisecond pulsars with known $r'$ and $i'$-band magnitudes are presented for comparison. 
These include PSR J0348+0432 \citep{antoniadis}, 
PSR J1012+5307 \citep{nicastro}, PSR J2317+1439 \citep{dai} and PSRs J0614$-$3329 and J1231$-$1411 
\citep{bassa}. The magnitudes of the sources were adopted both 
from the Sloan Digital Sky Survey (SDSS) archive \citep{abolfahti} and directly from the cited articles. 
The respective extinction correction values were calculated using the reddening models by \citet{green} and \citet{drimmel}. The 
dereddened magnitudes were then transformed into the absolute magnitudes using the DM distances and, if provided, the model-predicted WD distances 
and the parallax distances. 
The absolute magnitude and 
reddening for PSR J2317+1439 from \citet{dai} were recalculated based on the updated parallax information provided by \citet{arzoumanian}. 

As seen from the Fig.~\ref{fig:2}, the pulsar companion is in a good agreement with the cooling predictions and could 
represent a WD with a hydrogen or helium atmosphere and 
a mass of 0.2--0.7 M$_{\odot}$. 
According to the location on the colour-magnitude and colour-colour diagrams (see Fig.~\ref{fig:2} and Fig.~\ref{fig:3}), 
it is expected to have the effective temperature of $\sim$~6000 K and the cooling age of 1--2 Gyr, as follows from the cooling models. 
The $r'-i'$ colour does not exclude the possibility that the WD    
can fall onto the DA cooling sequence in the upper part of the colour-colour diagram, where the collision-induced 
absorption of molecular hydrogen strongly enhances the opacity in the 
infrared and shifts the spectral energy distribution to bluer colour indices (see, e.g., \citet{hansen}). 
In this case the expected effective temperature of the WD would be $\lesssim$~3000 K. However, the corresponding $r'$-band absolute magnitude would then  
imply a distance of $\lesssim$~230 pc. Such inconsistency with the distance derived from the parallax measurements \citep{arzoumanian} rejects this possibility.

To set constraints on the WD mass and effective temperature, we followed the method described by \citet{dai}. We obtained a single 
WD cooling model by unifying the cooling predictions for WD masses of 
0.19--0.25 M$_{\odot}$ from \citet{panei} and 0.3--1.2 M$_{\odot}$ from \citet{holberg}, \citet{kowalski}, \citet{tremblay} and \citet{bergeron2011}. 
We interpolated this model on the mass-temperature plane within the ranges of 
0.19--1.2 M$_{\odot}$ and 4000--11000 K using a 7000$\times$7000 grid. 
For each point of the plane we then calculated the likelihood using Equation (5) from \citet{dai}. 
In Fig.~\ref{fig:4} we present the mass-temperature plane with the resulting constraints on the WD mass and 
effective temperature 
and the respective 1D likelihoods with calculated median values of 0.52$^{+0.25}_{-0.19}$ \msun~and 6300$^{+1000}_{-800}$ K.  

\begin{figure}
\begin{center}
\begin{picture}(125,250)
\put (-75,65) {
\includegraphics[width=95mm]{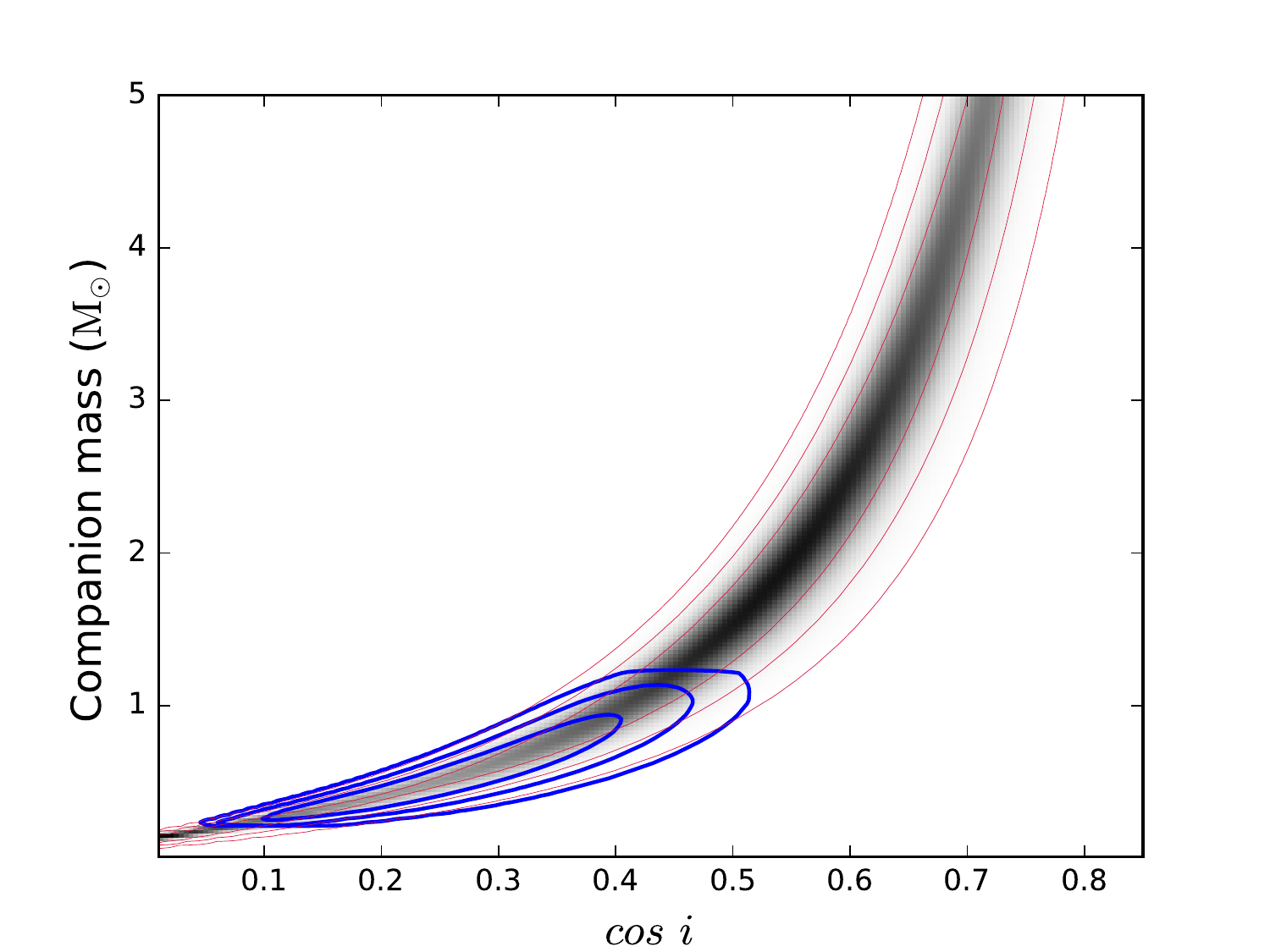}}
\put (-75,-140) {
\includegraphics[width=95mm]{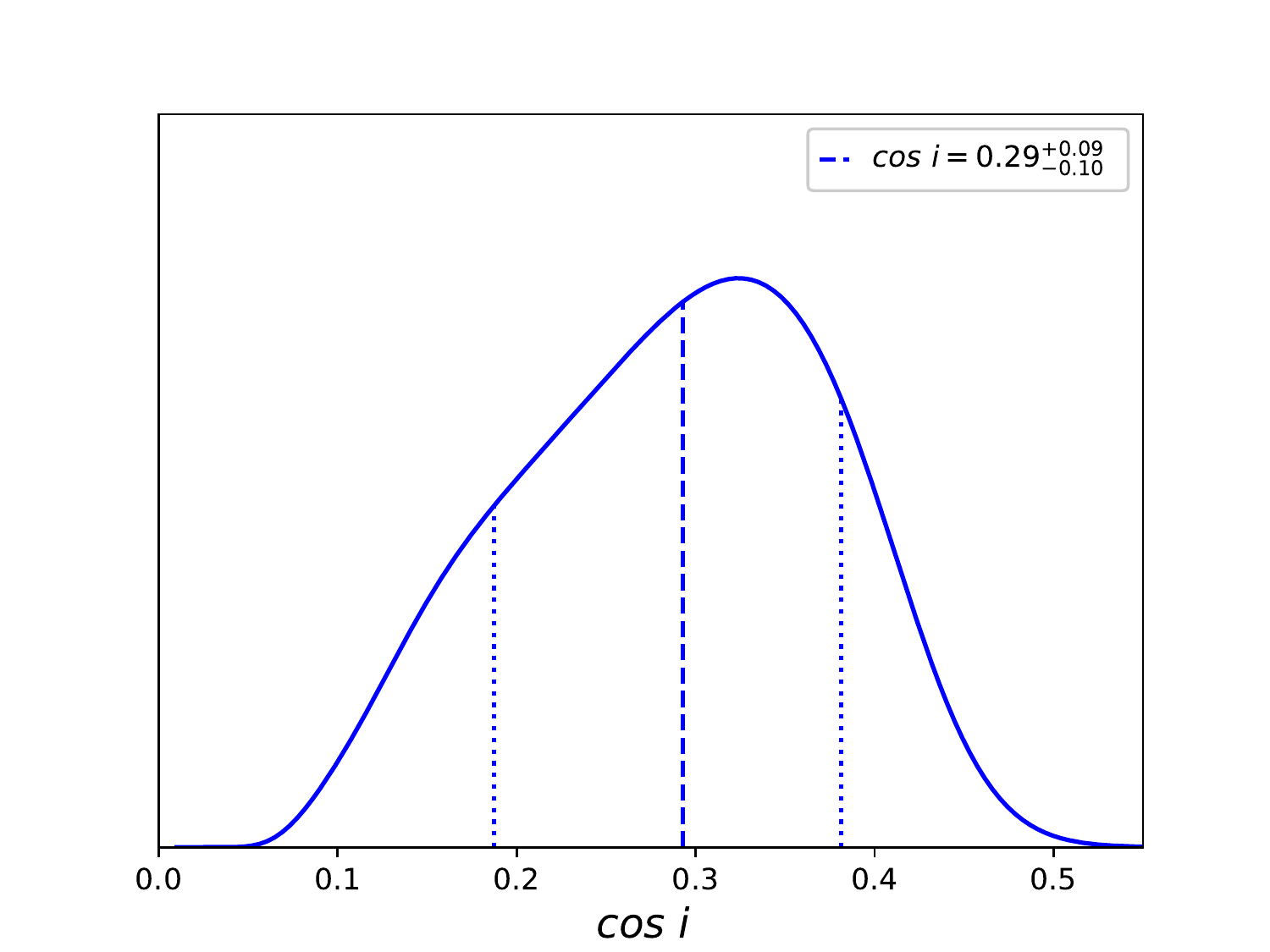}}
\end{picture}
\end{center}
\vspace{48mm}
\caption{Upper panel: constraints on the inclination angle and companion mass as derived by \citet{fonseca} from the radio timing alone (red contours) and  
after combining the radio timing and photometric analysis (blue contours). 
Contours represent 1$\sigma$, 2$\sigma$ and 3$\sigma$ confidence levels. Lower panel: the resulting 1D likelihood for the inclination angle. 
Dashed and dotted lines correspond to the median value and 1$\sigma$ confidence levels, respectively.}
\label{fig:5}
\end{figure}

To check whether our companion mass estimation can provide better constraints on other parameters of the pulsar system, we have used the probabilities of 
the pulsar Shapiro-delay parameters obtained from the NANOGrav nine-year dataset \citep{fonseca}. 
The Shapiro delay for the \psr~system ranks among the weakest detections in the set, and only rough constraints on the companion mass and 
inclination angle of $m_c=2.3^{+1.7}_{-1.3}$ M$_{\odot}$ and $i=54^{+12}_{-7}$ degrees were provided (see Fig.~\ref{fig:5}). We have combined our likelihoods on the 
companion mass with the probability map from \citet{fonseca}. The resulting constraints on the two parameters are shown in Fig.~\ref{fig:5}. 
Based on the new probability map, we have calculated the median value for the inclination angle of $i=73^{+6}_{-5}$ degrees, where the errors 
correspond to 1$\sigma$ uncertainties. 
Using the binary mass function and the resulting constraint on the inclination angle, we have then calculated the pulsar mass value of $m_p=3.1^{+2.7}_{-2.0}$ M$_{\odot}$.

\section{Summary and discussion}
\label{sec:sum}

The first deep optical observations of the \psr~field with the GTC/OSIRIS 
have allowed us to reveal a point source whose position coincides with the pulsar radio timing position and whose magnitudes are in agreement 
with a WD. We propose the source 
as a likely optical identification of the pulsar companion. 

We have considered that the optical source can be related to stellar families other than the WDs. Based on its effective temperature, the source could belong 
to the subdwarf branch. However, the subdwarf absolute magnitudes and the object apparent magnitude would then imply an unreasonably high distance of $\sim$30 kpc, 
placing it outside the galaxy. Its association with the main sequence would assign it even larger distance, thus ruling out any stellar nature other than  
WD. Although the extragalactic origin cannot be excluded based on the current information, it is very unlikely that the source represents 
a distant galaxy. These conclusions favour a true detection of the pulsar companion. 

Assuming the companion is indeed a WD, 
we have put constrains on its mass of 0.52$^{+0.25}_{-0.19}$ \msun~and temperature of 6300$^{+1000}_{-800}$ K using the WD evolutionary tracks. 
The derived parameters imply that the companion belongs to 
the population of He or CO-core WDs. 
By the orbital period $P_b$, the pulsar period $P_s$ and the companion mass $m_c$, the \psr~system is found to fit in well the $P_b$--$m_c$ and $P_s$--$m_c$
distributions of the known binary MSPs (see, e.g., Fig.8 of \citet{manchester_mil}). It resides among other systems with a similar companion mass, $P_s$ $\sim$ 4--6 ms   
and $P_b$ $\sim$~30--300 days. On the other hand, considering the \psr~companion represents a He-core WD, its mass can be predicted based on the ($P_b$, $M_c$) correlation
introduced by \citet{tauris1999}. The correlation is estimated using the models of low-mass X-ray binaries and in case of our $P_b$ gives the range of 
0.34--0.37 \msun, depending on the chemical 
composition of the donor star. Our constraint on the WD mass is consistent with this prediction, 
supporting a genuine association with the \psr~system. 

Combining our estimation on the companion mass with the radio timing measurements, we have obtained new constraints on the binary system inclination angle of 
$i=73^{+6}_{-5}$ degrees and the pulsar mass of $m_p=3.1^{+2.7}_{-2.0}$ M$_{\odot}$. 
The latter represents a more accurate 
constraint on the pulsar mass as compared to the radio timing estimations alone and it is also consistent with the range of 1.37--2.01 M$_{\odot}$ found in  
precise mass measurements of recycled pulsars \citep{ozel}.  

In order to reveal the fundamental parameters of the companion, photometry in other bands, optical spectroscopy and parallax 
measurements are needed. 
Spectroscopic observations will allow one to set more stringent constraints on its atmospheric parameters, temperature and surface gravity.   
Despite the fact that the source is relatively faint, optical spectroscopy is still feasible with the 8--10 meter class telescopes. 
Combined together with the 
precise distance measurements, it can potentially lead to the mass determination both for the companion and the neutron star. 

\section*{Acknowledgements}

The authors thank the anonymous referee for useful comments which allowed them to improve the manuscript, Emmanuel Fonseca (on behalf of the NANOGrav collaboration) 
for providing probabilities of the Shapiro-delay parameters from the NANOGrav nine-year data set 
and Maxim Voronkov for helpful discussions. The work of AVK and DAZ (Sect.~\ref{sec:res}) was supported by the Russian Foundation for Basic
Research, project No. 18-32-00781 mol\_a. AYuK and SVZ acknowledge PAPIIT grant IN-100617 for resources provided towards this research. 
The manuscript is based on observations made with the Gran Telescopio Canarias (GTC), installed in the Spanish Observatorio del Roque de los Muchachos 
of the Instituto de Astrof\'isica de Canarias, 
in the island of La Palma. 
IRAF is distributed by the National Optical Astronomy Observatory, which is operated by the Association of Universities for Research in 
Astronomy (AURA) under a cooperative agreement with the National Science Foundation. 
This research has made use of the USNOFS Image and Catalogue Archive operated by the United States Naval Observatory, Flagstaff Station 
(http://www.nofs.navy.mil/data/fchpix/). Funding for the Sloan Digital Sky Survey IV has been provided by the Alfred P. Sloan Foundation, 
the U.S. Department of Energy Office of Science, and the Participating Institutions. SDSS-IV acknowledges
support and resources from the Center for High-Performance Computing at
the University of Utah. The SDSS web site is www.sdss.org. 
The authors used WD colour tables available at http://www.astro.umontreal.ca/~bergeron/CoolingModels.






\bibliographystyle{mnras}
\bibliography{2302}

\begin{thebibliography}{}
\makeatletter
\relax
\def\mn@urlcharsother{\let\do\@makeother \do\$\do\&\do\#\do\^\do\_\do\%\do\~}
\def\mn@doi{\begingroup\mn@urlcharsother \@ifnextchar [ {\mn@doi@}
  {\mn@doi@[]}}
\def\mn@doi@[#1]#2{\def\@tempa{#1}\ifx\@tempa\@empty \href
  {http://dx.doi.org/#2} {doi:#2}\else \href {http://dx.doi.org/#2} {#1}\fi
  \endgroup}
\def\mn@eprint#1#2{\mn@eprint@#1:#2::\@nil}
\def\mn@eprint@arXiv#1{\href {http://arxiv.org/abs/#1} {{\tt arXiv:#1}}}
\def\mn@eprint@dblp#1{\href {http://dblp.uni-trier.de/rec/bibtex/#1.xml}
  {dblp:#1}}
\def\mn@eprint@#1:#2:#3:#4\@nil{\def\@tempa {#1}\def\@tempb {#2}\def\@tempc
  {#3}\ifx \@tempc \@empty \let \@tempc \@tempb \let \@tempb \@tempa \fi \ifx
  \@tempb \@empty \def\@tempb {arXiv}\fi \@ifundefined
  {mn@eprint@\@tempb}{\@tempb:\@tempc}{\expandafter \expandafter \csname
  mn@eprint@\@tempb\endcsname \expandafter{\@tempc}}}

\bibitem[\protect\citeauthoryear{{Abolfathi} et~al.,}{{Abolfathi}
  et~al.}{2017}]{abolfahti}
{Abolfathi} B.,  et~al., 2017, preprint, \href
  {http://adsabs.harvard.edu/abs/2017arXiv170709322A} {} (\mn@eprint {arXiv}
  {1707.09322})

\bibitem[\protect\citeauthoryear{{Alpar}, {Cheng}, {Ruderman}  \&
  {Shaham}}{{Alpar} et~al.}{1982}]{alpar}
{Alpar} M.~A.,  {Cheng} A.~F.,  {Ruderman} M.~A.,   {Shaham} J.,  1982, \mn@doi
  [\nat] {10.1038/300728a0}, \href
  {http://adsabs.harvard.edu/abs/1982Natur.300..728A} {300, 728}

\bibitem[\protect\citeauthoryear{{Antoniadis} et~al.,}{{Antoniadis}
  et~al.}{2013}]{antoniadis}
{Antoniadis} J.,  et~al., 2013, \mn@doi [Science] {10.1126/science.1233232},
  \href {http://adsabs.harvard.edu/abs/2013Sci...340..448A} {340, 448}

\bibitem[\protect\citeauthoryear{{Arzoumanian} et~al.,}{{Arzoumanian}
  et~al.}{2018}]{arzoumanian}
{Arzoumanian} Z.,  et~al., 2018, preprint, \href
  {http://adsabs.harvard.edu/abs/2018arXiv180101837A} {} (\mn@eprint {arXiv}
  {1801.01837})

\bibitem[\protect\citeauthoryear{{Bassa}, {Antoniadis}, {Camilo}, {Cognard},
  {Koester}, {Kramer}, {Ransom}  \& {Stappers}}{{Bassa} et~al.}{2016}]{bassa}
{Bassa} C.~G.,  {Antoniadis} J.,  {Camilo} F.,  {Cognard} I.,  {Koester} D.,
  {Kramer} M.,  {Ransom} S.~R.,   {Stappers} B.~W.,  2016, \mn@doi [\mnras]
  {10.1093/mnras/stv2607}, \href
  {http://adsabs.harvard.edu/abs/2016MNRAS.455.3806B} {455, 3806}

\bibitem[\protect\citeauthoryear{{Bergeron} et~al.,}{{Bergeron}
  et~al.}{2011}]{bergeron2011}
{Bergeron} P.,  et~al., 2011, \mn@doi [\apj] {10.1088/0004-637X/737/1/28},
  \href {http://adsabs.harvard.edu/abs/2011ApJ...737...28B} {737, 28}

\bibitem[\protect\citeauthoryear{{Bisnovatyi-Kogan} \&
  {Komberg}}{{Bisnovatyi-Kogan} \& {Komberg}}{1974}]{kogan}
{Bisnovatyi-Kogan} G.~S.,  {Komberg} B.~V.,  1974, \sovast, \href
  {http://adsabs.harvard.edu/abs/1974SvA....18..217B} {18, 217}

\bibitem[\protect\citeauthoryear{{Cognard} et~al.,}{{Cognard}
  et~al.}{2011}]{cognard}
{Cognard} I.,  et~al., 2011, \mn@doi [\apj] {10.1088/0004-637X/732/1/47}, \href
  {http://adsabs.harvard.edu/abs/2011ApJ...732...47C} {732, 47}

\bibitem[\protect\citeauthoryear{{Cordes} \& {Lazio}}{{Cordes} \&
  {Lazio}}{2002}]{cordes}
{Cordes} J.~M.,  {Lazio} T.~J.~W.,  2002, ArXiv Astrophysics e-prints, \href
  {http://adsabs.harvard.edu/abs/2002astro.ph..7156C} {}

\bibitem[\protect\citeauthoryear{{Dai}, {Smith}, {Wang}, {Okamoto}, {Xu}, {Yue}
   \& {Liu}}{{Dai} et~al.}{2017}]{dai}
{Dai} S.,  {Smith} M.~C.,  {Wang} S.,  {Okamoto} S.,  {Xu} R.~X.,  {Yue} Y.~L.,
    {Liu} J.~F.,  2017, \mn@doi [\apj] {10.3847/1538-4357/aa7209}, \href
  {http://adsabs.harvard.edu/abs/2017ApJ...842..105D} {842, 105}

\bibitem[\protect\citeauthoryear{{Drimmel}, {Cabrera-Lavers}  \&
  {L{\'o}pez-Corredoira}}{{Drimmel} et~al.}{2003}]{drimmel}
{Drimmel} R.,  {Cabrera-Lavers} A.,   {L{\'o}pez-Corredoira} M.,  2003, \mn@doi
  [\aap] {10.1051/0004-6361:20031070}, \href
  {http://adsabs.harvard.edu/abs/2003A%26A...409..205D} {409, 205}

\bibitem[\protect\citeauthoryear{{Fonseca} et~al.,}{{Fonseca}
  et~al.}{2016}]{fonseca}
{Fonseca} E.,  et~al., 2016, \mn@doi [\apj] {10.3847/0004-637X/832/2/167},
  \href {http://adsabs.harvard.edu/abs/2016ApJ...832..167F} {832, 167}

\bibitem[\protect\citeauthoryear{{Green} et~al.,}{{Green} et~al.}{2015}]{green}
{Green} G.~M.,  et~al., 2015, \mn@doi [\apj] {10.1088/0004-637X/810/1/25},
  \href {http://adsabs.harvard.edu/abs/2015ApJ...810...25G} {810, 25}

\bibitem[\protect\citeauthoryear{{Guillemot} et~al.,}{{Guillemot}
  et~al.}{2016}]{guillemot}
{Guillemot} L.,  et~al., 2016, \mn@doi [\aap] {10.1051/0004-6361/201527847},
  \href {http://adsabs.harvard.edu/abs/2016A%26A...587A.109G} {587, A109}

\bibitem[\protect\citeauthoryear{{Hansen}}{{Hansen}}{1998}]{hansen}
{Hansen} B.~M.~S.,  1998, \mn@doi [\nat] {10.1038/29710}, \href
  {http://adsabs.harvard.edu/abs/1998Natur.394..860H} {394, 860}

\bibitem[\protect\citeauthoryear{{Holberg} \& {Bergeron}}{{Holberg} \&
  {Bergeron}}{2006}]{holberg}
{Holberg} J.~B.,  {Bergeron} P.,  2006, \mn@doi [\aj] {10.1086/505938}, \href
  {http://adsabs.harvard.edu/abs/2006AJ....132.1221H} {132, 1221}

\bibitem[\protect\citeauthoryear{{Kowalski} \& {Saumon}}{{Kowalski} \&
  {Saumon}}{2006}]{kowalski}
{Kowalski} P.~M.,  {Saumon} D.,  2006, \mn@doi [\apjl] {10.1086/509723}, \href
  {http://adsabs.harvard.edu/abs/2006ApJ...651L.137K} {651, L137}

\bibitem[\protect\citeauthoryear{{Manchester}}{{Manchester}}{2017}]{manchester_mil}
{Manchester} R.~N.,  2017, \mn@doi [Journal of Astrophysics and Astronomy]
  {10.1007/s12036-017-9469-2}, \href
  {http://adsabs.harvard.edu/abs/2017JApA...38...42M} {38, 42}

\bibitem[\protect\citeauthoryear{{Manchester}, {Hobbs}, {Teoh}  \&
  {Hobbs}}{{Manchester} et~al.}{2005}]{manchester}
{Manchester} R.~N.,  {Hobbs} G.~B.,  {Teoh} A.,   {Hobbs} M.,  2005, \mn@doi
  [\aj] {10.1086/428488}, \href
  {http://adsabs.harvard.edu/abs/2005AJ....129.1993M} {129, 1993}

\bibitem[\protect\citeauthoryear{{Matthews} et~al.,}{{Matthews}
  et~al.}{2016}]{matthews}
{Matthews} A.~M.,  et~al., 2016, \mn@doi [\apj] {10.3847/0004-637X/818/1/92},
  \href {http://adsabs.harvard.edu/abs/2016ApJ...818...92M} {818, 92}

\bibitem[\protect\citeauthoryear{{Nicastro}, {Lyne}, {Lorimer}, {Harrison},
  {Bailes}  \& {Skidmore}}{{Nicastro} et~al.}{1995}]{nicastro}
{Nicastro} L.,  {Lyne} A.~G.,  {Lorimer} D.~R.,  {Harrison} P.~A.,  {Bailes}
  M.,   {Skidmore} B.~D.,  1995, \mn@doi [\mnras] {10.1093/mnras/273.1.L68},
  \href {http://adsabs.harvard.edu/abs/1995MNRAS.273L..68N} {273, L68}

\bibitem[\protect\citeauthoryear{{{\"O}zel} \& {Freire}}{{{\"O}zel} \&
  {Freire}}{2016}]{ozel}
{{\"O}zel} F.,  {Freire} P.,  2016, \mn@doi [\araa]
  {10.1146/annurev-astro-081915-023322}, \href
  {http://adsabs.harvard.edu/abs/2016ARA%26A..54..401O} {54, 401}

\bibitem[\protect\citeauthoryear{{Panei}, {Althaus}, {Chen}  \& {Han}}{{Panei}
  et~al.}{2007}]{panei}
{Panei} J.~A.,  {Althaus} L.~G.,  {Chen} X.,   {Han} Z.,  2007, \mn@doi
  [\mnras] {10.1111/j.1365-2966.2007.12400.x}, \href
  {http://adsabs.harvard.edu/abs/2007MNRAS.382..779P} {382, 779}

\bibitem[\protect\citeauthoryear{{Ray} et~al.,}{{Ray} et~al.}{2012}]{ray}
{Ray} P.~S.,  et~al., 2012, preprint, \href
  {http://adsabs.harvard.edu/abs/2012arXiv1205.3089R} {} (\mn@eprint {arXiv}
  {1205.3089})

\bibitem[\protect\citeauthoryear{{Saz Parkinson} \& {Fermi LAT
  Collaboration}}{{Saz Parkinson} \& {Fermi LAT Collaboration}}{2013}]{saz}
{Saz Parkinson} P.~M.,  {Fermi LAT Collaboration} 2013, in {van Leeuwen} J.,
  ed.,  IAU Symposium Vol. 291, Neutron Stars and Pulsars: Challenges and
  Opportunities after 80 years. pp 81--86 (\mn@eprint {arXiv} {1210.7530}),
  \mn@doi{10.1017/S174392131202323X}

\bibitem[\protect\citeauthoryear{{Schlafly} \& {Finkbeiner}}{{Schlafly} \&
  {Finkbeiner}}{2011}]{schlafly}
{Schlafly} E.~F.,  {Finkbeiner} D.~P.,  2011, \mn@doi [\apj]
  {10.1088/0004-637X/737/2/103}, \href
  {http://adsabs.harvard.edu/abs/2011ApJ...737..103S} {737, 103}

\bibitem[\protect\citeauthoryear{{Shapiro}}{{Shapiro}}{1964}]{shapiro}
{Shapiro} I.~I.,  1964, \mn@doi [Physical Review Letters]
  {10.1103/PhysRevLett.13.789}, \href
  {http://adsabs.harvard.edu/abs/1964PhRvL..13..789S} {13, 789}

\bibitem[\protect\citeauthoryear{{Smith} et~al.,}{{Smith} et~al.}{2002}]{smith}
{Smith} J.~A.,  et~al., 2002, \mn@doi [\aj] {10.1086/339311}, \href
  {http://adsabs.harvard.edu/abs/2002AJ....123.2121S} {123, 2121}

\bibitem[\protect\citeauthoryear{{Tauris} \& {Savonije}}{{Tauris} \&
  {Savonije}}{1999}]{tauris1999}
{Tauris} T.~M.,  {Savonije} G.~J.,  1999, \aap, \href
  {http://adsabs.harvard.edu/abs/1999A%26A...350..928T} {350, 928}

\bibitem[\protect\citeauthoryear{{Tauris}, {Langer}  \& {Kramer}}{{Tauris}
  et~al.}{2011}]{tauris2011}
{Tauris} T.~M.,  {Langer} N.,   {Kramer} M.,  2011, \mn@doi [\mnras]
  {10.1111/j.1365-2966.2011.19189.x}, \href
  {http://adsabs.harvard.edu/abs/2011MNRAS.416.2130T} {416, 2130}

\bibitem[\protect\citeauthoryear{{Tremblay}, {Bergeron}  \&
  {Gianninas}}{{Tremblay} et~al.}{2011}]{tremblay}
{Tremblay} P.-E.,  {Bergeron} P.,   {Gianninas} A.,  2011, \mn@doi [\apj]
  {10.1088/0004-637X/730/2/128}, \href
  {http://adsabs.harvard.edu/abs/2011ApJ...730..128T} {730, 128}

\bibitem[\protect\citeauthoryear{{Yao}, {Manchester}  \& {Wang}}{{Yao}
  et~al.}{2017}]{yao}
{Yao} J.~M.,  {Manchester} R.~N.,   {Wang} N.,  2017, \mn@doi [\apj]
  {10.3847/1538-4357/835/1/29}, \href
  {http://adsabs.harvard.edu/abs/2017ApJ...835...29Y} {835, 29}

\makeatother
\end{thebibliography}








\bsp	
\label{lastpage}
\end{document}